\documentclass[conference]{IEEEtran}
\IEEEoverridecommandlockouts

\usepackage{cite}
\usepackage{amsmath,amssymb,amsfonts}
\usepackage[linesnumbered,ruled,vlined]{algorithm2e}
\usepackage{algpseudocode}
\usepackage{graphicx}
\graphicspath{{figs/}}
\usepackage{textcomp}
\usepackage{xcolor}
\usepackage{caption}
\def\BibTeX{{\rm B\kern-.05em{\sc i\kern-.025em b}\kern-.08em
    T\kern-.1667em\lower.7ex\hbox{E}\kern-.125emX}}
\begin{document}

\title{Angular Velocity Estimation of Image Motion Mimicking the Honeybee Tunnel Centring Behaviour \\
\thanks{ This research is funded by the EU HORIZON 2020 project, STEP2DYNA (grant agreement No. 691154) and ULTRACEPT (grant agreement No. 778062); the National Natural Science Foundation of China (grant agreement No. 11771347)}
}

\author{
	\IEEEauthorblockN{Huatian Wang\IEEEauthorrefmark{1},   Qinbing Fu\IEEEauthorrefmark{3}\IEEEauthorrefmark{4}, Hongxin Wang\IEEEauthorrefmark{1}, Jigen Peng\IEEEauthorrefmark{2}\IEEEauthorrefmark{4}, Paul Baxter\IEEEauthorrefmark{1}, Cheng Hu\IEEEauthorrefmark{3}\IEEEauthorrefmark{4}, Shigang Yue\IEEEauthorrefmark{1}\IEEEauthorrefmark{3}\IEEEauthorrefmark{4}}
	
	\IEEEauthorblockA{\IEEEauthorrefmark{1} School of Computer Science, University of Lincoln, Lincoln, UK}
	\IEEEauthorblockA{\IEEEauthorrefmark{2} School of Mathematics and Information Science, Guangzhou University, Guangzhou, China}
	\IEEEauthorblockA{\IEEEauthorrefmark{3} School of Mechanical and Electrical Engineering, Guangzhou University, Guangzhou, China}
    \IEEEauthorblockA{\IEEEauthorrefmark{4} Machine Life and Intelligent Research Centre, Guangzhou University, Guangzhou, China}
	\{hwang, qifu, howang\}@lincoln.ac.uk, jgpeng@gzhu.edu.cn, \{pbaxter, chu, syue\}@lincoln.ac.uk
}

\maketitle

\begin{abstract}
Insects use visual information to estimate angular velocity of retinal image motion, which determines a variety of flight behaviours including speed regulation, tunnel centring and visual navigation. For angular velocity estimation, honeybees show large spatial-independence against visual stimuli, whereas the previous models have not fulfilled such an ability. To address this issue, we propose a bio-plausible model for estimating the image motion velocity based on behavioural experiments of the honeybee flying through patterned tunnels. The proposed model contains mainly three parts, the texture estimation layer for spatial information extraction, the delay-and-correlate layer for temporal information extraction and the decoding layer for angular velocity estimation. This model produces responses that are largely independent of the spatial frequency in grating experiments. And the model has been implemented in a virtual bee for tunnel centring simulations. The results coincide with both electro-physiological neuron spike and behavioural path recordings, which indicates our proposed method provides a better explanation of the honeybee's image motion detection mechanism guiding the tunnel centring behaviour.
\end{abstract}


\section{Introduction}

It has been researched for a long time to understand how insects, like honeybees and locusts, detect image motion and even determine its direction and angular velocity (degrees passed per second of the image motion across the retina)\cite{yue2006collision, FU2018127, wang2018directionally}. The underlying neural mechanism is essential for explaining a variety of the behaviours including course maintaining\cite{Sri1997}, distance estimation\cite{Esch1995} and speed regulation\cite{Sri2000}. Lots of the honeybee's behavioural experiments have been conducted due to its excellent flight controlling ability. For instance, in the tunnel centring experiments, it is likely that honeybees fly along the central route of the tunnel by balancing the angular velocities sensed on both eyes\cite{Sri1996}. Further researches revealed that in the honeybee's central nerve cord, the responses of some descending neurons grow as the angular velocity increases \cite{Ibb2001}\cite{Ibb2017}. Both indicate that the honeybee can estimate the angular velocity of image motion. However, the neural mechanism behind this ability still has not been fully understood.

\begin{figure}[htb]
\centering
\includegraphics [width=70mm]{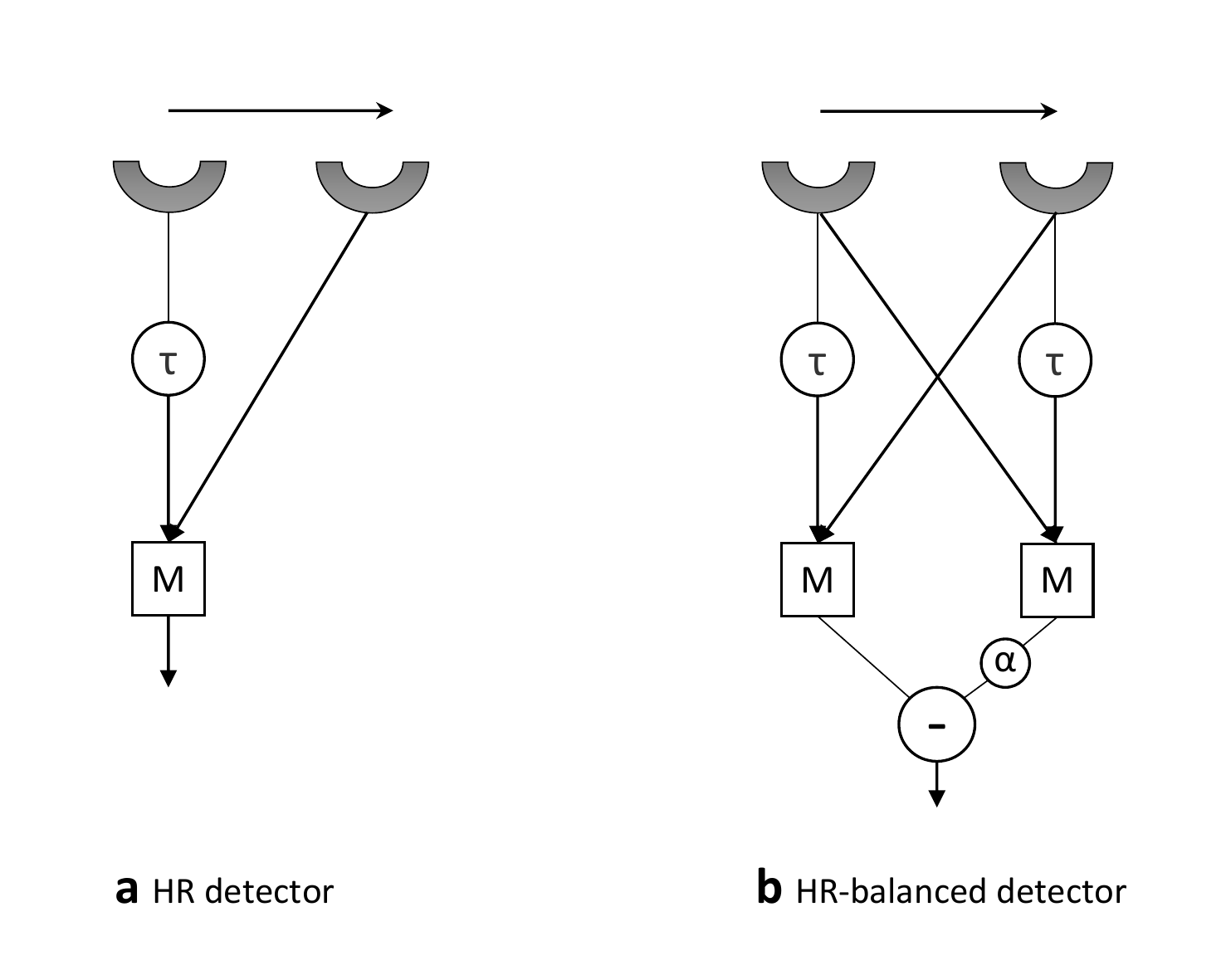}
\captionsetup{font={small}, name={Fig.}}
\caption{HR detector and HR-balanced detector. (a) In the Hassenstein-Reichardt detector, a delayed signal from left photoreceptor multiplies (M) the signal from right to give a preferred direction enhancement response \cite{HR1956}. (b) HR-balanced detector uses balance parameter $\alpha$ to tune the dependence on spatial frequency \cite{Zanker1999}.}
\label{fig1}
\end{figure}

Back to the last century, Hassenstein and Richardt proposed a classic elementary motion detecting model describing the mechanism of motion sensing\cite{HR1956}. This so-called HR model uses visual signals from two neighbouring viewpoints to detect a preferred motion. The signal from left arm is delayed and then multiplies the non-delay signal from right arm to get a directional response which is much higher when the motion is progressive (see Fig. \ref{fig1}(a)).

The HR model with such a 'delay-and-correlate' mechanism derives many variations including the HR-balanced model\cite{Zanker1999}, which has a symmetrical structure and is suitable for sensing motion along both preferred and opposite directions. The balance parameter $\alpha$ (see Fig. \ref{fig1}(b)) on the regressive arm between 0 and 1 can tune the spatial frequency dependence of the finally response according to the numerical experiments. However, both the HR model and the HR-balanced model are tuned for particular temporal frequency rather than angular velocity. Zanker et al.\cite{Zanker1999} suggest that ratio of two HR-balanced detectors with different optimal temporal frequencies may provide an angular velocity tuned response.

Based on this idea, Cope et al. \cite{Cope2016} built up an angular velocity model, C-HR model, using the ratio of two HR-balanced detectors with different time delays. And the response of the model is basically in accordance with the spike recordings from Ibbotson's work on electro-physiological experiments \cite{Ibb2001} and largely independent of the spatial frequency of the moving pattern. However, this model only performs well at the spatial frequency of around 100 $^\circ/s$. This is different from that the honeybees usually maintain a constant angular velocity around 300 $^\circ/s$ in an open flight\cite{Baird2005}. And the detector does not produce a relatively higher response at low temporal frequencies as the experimental data show, which suggests that the real neural implementation may be different from their model. Riabinina and Philippides\cite{Ria2009} also built up a model, R-HR model, with three input viewpoints for estimation of the angular velocity. The response of their model slightly depend on the spatial frequency of the moving grating. And Cope et al.\cite{Cope2016} argued that R-HR model is energy inefficient since it separates angular velocity estimation circuit from the optomotor circuit which requires much more additional neurons. Wang et al.\cite{Wang2018} proposed a model based on the neural structure of Drosophila's visual system, which compares the results from detectors with different sampling rates, to get a spatial independent response. However, the independence still needs to be more significant before reproducing the centring behaviours of the honeybees in simulations.

Previous mentioned models estimate the angular velocity directly without considering an estimation of the spatial frequency. However their model structure (based on HR model) determines that the responses will depend on the spatial frequency more or less from the very beginning. Considering the characteristics of the insects' compound eyes which have thousands of ommatidia and much higher temporal resolution, it is probably a better choice to estimate angular velocity by combining the spatial and temporal information together.

The ommatidia are arranged hexagonally and each corresponds to a visual column in the visual system. So it is possible to get the spatial frequency of the image roughly by sensing the light intensities received by different ommatidia. At the same time, the temporal frequency information can be extracted from the HR-balanced model. Based on the behavioural experiments of the honeybees mentioned above, we will present a new model combining both the spatial and temporal information of the input visual signals to estimate the angular velocity of the image motion more accurately, which provides a possible explanation of the neural mechanism beneath the angular velocity detecting circuits.

The rest of this paper is organised as follows: Section II presents the formulation of the proposed angular velocity decoding model (AVDM). Sections III exhibits the experiments and results. Finally we conclude this research and give further discussion in Section IV.

\section{Method}

\subsection{Input Signals Simulation}

\begin{figure}[htb]
\centering
\includegraphics [width=88mm]{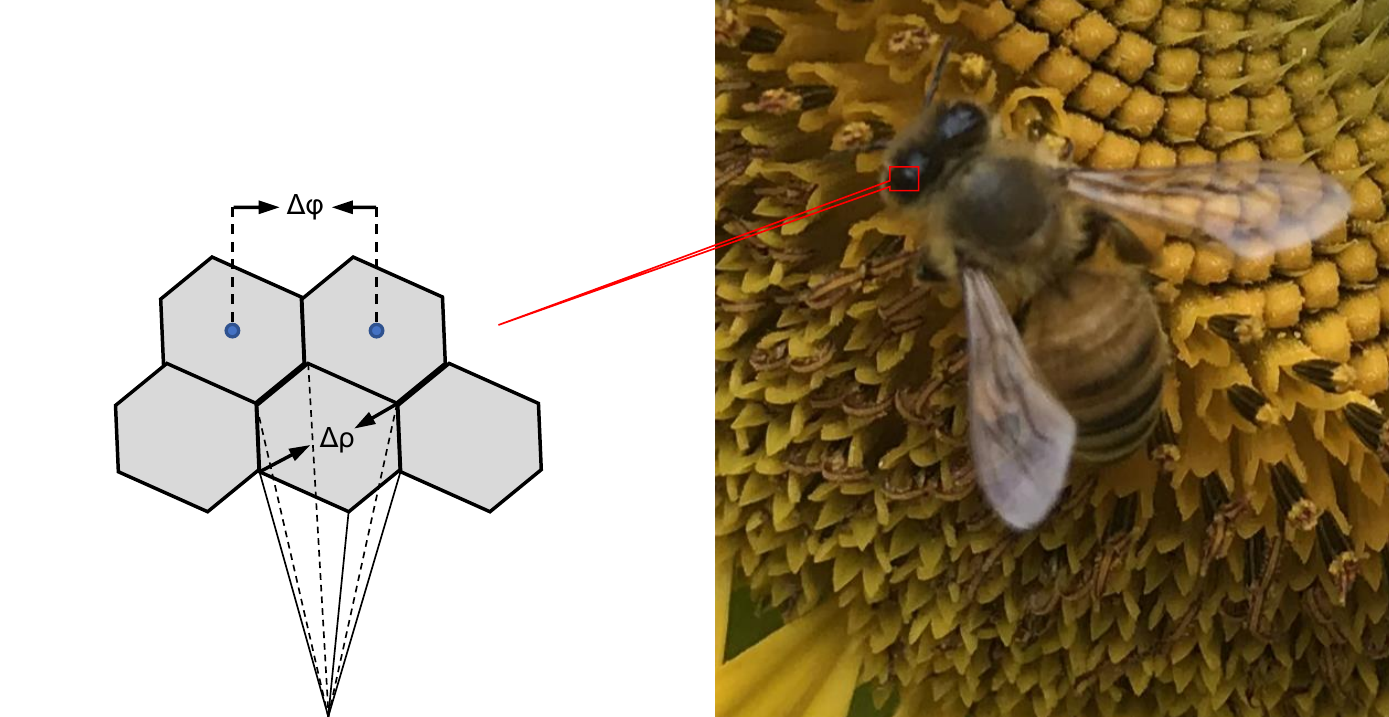}
\captionsetup{font={small},name={Fig.}}
\caption{Visual structures of the honeybee's compound eyes. The ommatidia are arranged hexagonally with a separation $\Delta \varphi$ (interommatidial angle) and each has its own small receptive field $\Delta \rho$ (acceptance angle).}
\label{fig2}
\end{figure}

In order to have a more bio-plausible parameter setting, first we need to investigate the spatial and temporal resolution of the honeybee. The spatial resolution of the honeybee's compound eye is mainly determined by two parameters: the interommatidial angle $\Delta \varphi$ and the acceptance angle $\Delta \rho$ (see Fig. \ref{fig2}). The optical axes of the neighboring ommatidia is separated by $\Delta \varphi$ (around 2 degrees), which varies in different regions \cite{Seidl1982}. And each ommatidium accepts light from a cone-shaped region with an acceptance angle $\Delta \rho$ (about 2.5 degrees) \cite{Laugh1971}. As for temporal frequency, though the critical fusion frequency (beyond which electroretinogram shows no response to the flickering light source) of the honeybee is 165-300 Hz \cite{Autrum1950}, the behavioural experiment which trains honeybees to distinguish rotating striped disk shows that honeybees can only resolve intensity fluctuation less than 200 Hz when the stimuli is moving \cite{Sri1984}. This helps us to build up a new angular velocity detecting model with a bio-plausible parameter setting when explaining how honeybees detect motion.

Here we simulate the input signals as two dimensional image frames of the sinusoidal gratings moving across the retina. We set the sample rate as 200 Hz, that is 5 ms per frame, which is in accordance with the honeybee's ability of the high temporal frequency processing. And let $\lambda$ and V be the spatial period ($^\circ$) and the moving speed ($^\circ/s$) of the grating movements, then the temporal frequency and angular frequency will be V/$\lambda$ and  $\omega = 2\pi V/\lambda$. Suppose the angular separation between pixels is $\varphi$, set to 2$^\circ$ in step with honeybee's spatial resolution, then the input images can be expressed as following:
\begin{equation}
I(x,y,t) = (\sin(\omega (t - \varphi  (y-1)/V))+1/C)/(1/C+1),
\label{eq1}
\end{equation}
where (x,y) denotes the location of the ommatidium, t indicates the time and $C \in (0,1]$ denotes the image contrast. For simplicity, we scale the pixel-wise intensity to be between 0 and 1 when the contrast is 1. Here we use Michelson contrast which is defined as the following:
\begin{equation}
C = \frac{I_{max}-I_{min}}{I_{max} + I_{min}},
\end{equation}
where the $I_{max}$ and $I_{min}$ ($I_{max}, I_{min} \geq 0$) indicate the highest and the lowest light intensities of the input signal.

\subsection{Angular Velocity Decoding Model}
\begin{figure}[htb]
\centering
\includegraphics [width=88mm]{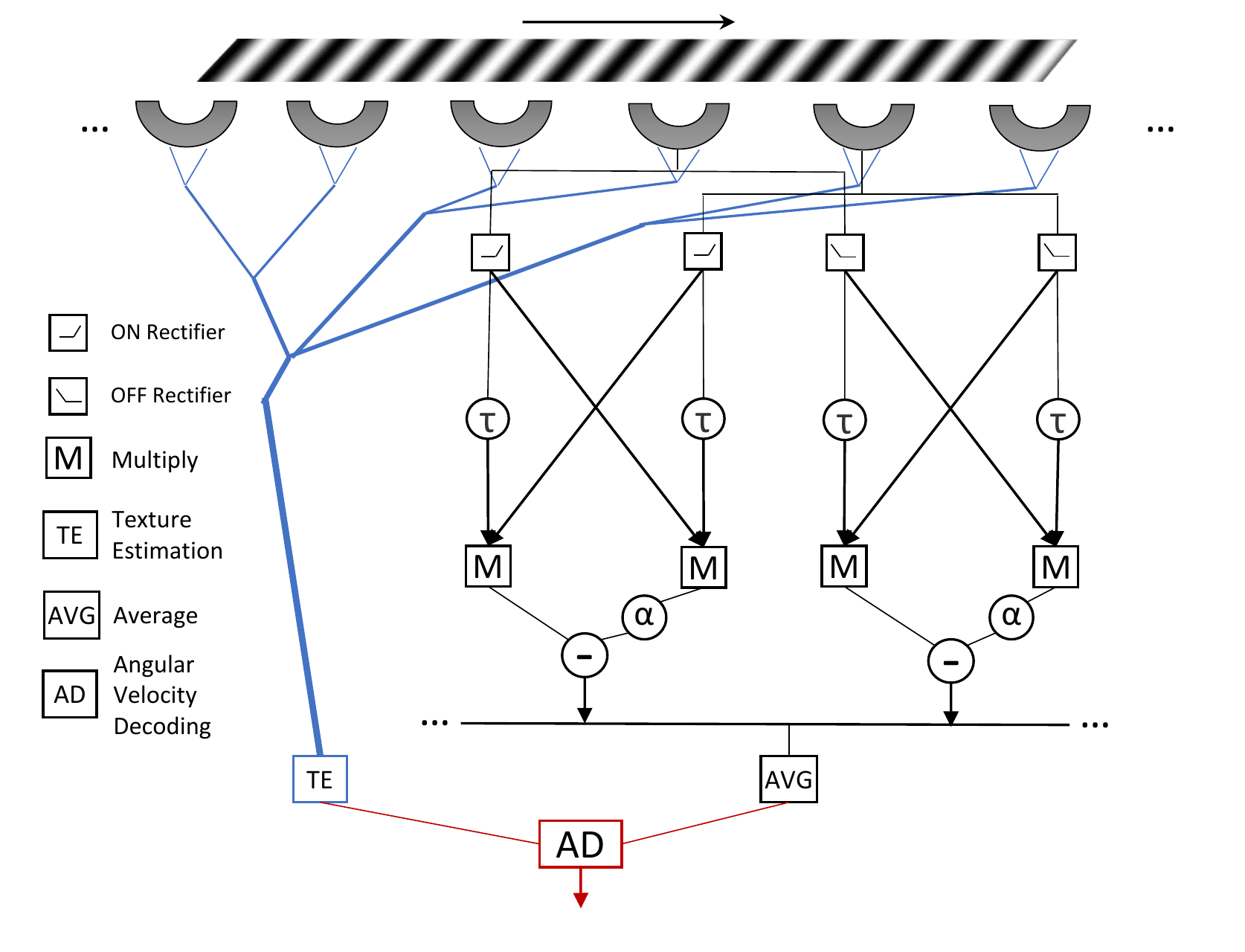}
\caption{ The structure of the proposed Angular Velocity Decoding Model. The visual information of the grating's moving is received by ommatidia. The global spatial frequency is estimated by texture estimation part, and the motion information is processed by motion detectors. In the proposed model, every motion detector receives the light intensity change from the neighbouring ommatidia. And the light intensity change is separated into ON and OFF pathways and then processed by two HR-balanced detectors. The texture information and the motion information from the average of the detectors across the whole vision field are combined. Then angular velocity is decoded from this composite information. }
\label{fig3}
\end{figure}

The model mainly contains 3 parts, the texture estimation layer for spatial information extraction, the delay and correlate layer for temporal information extraction and the decoding layer for angular velocity estimation. The structure of the proposed model is shown in Fig. \ref{fig3}.

\subsubsection{Texture estimation layer}

The simulated input signals received by retina are first processed by a texture estimation layer where the spatial frequency of the gratings is estimated by the light intensities of different locations. This is based on a hypothesis that insects can have a sense of the complexity of the texture. In fact, honeybees can discriminate patterns by visual cues including edge orientation, size and disruption \cite{roper2017insect}, which indicates that the assumption is reasonable. This texture estimation layer aims to get the spatial frequency with low computations.
 
Following the setting that every ommatidium covers 2$^\circ$ view \cite{Seidl1982}, with 66 horizontal by 60 vertical receptors per eye covering the view of 132$^\circ$ by 120$^\circ$, we can estimate the spatial period $\widehat{\lambda}$ of the gratings according to the light intensities. First the input image is transferred into a binary image by the relative intensity threshold $I_{thre} = (I_{max} - I_{min})/2$. Then the spatial frequency is estimated by counting the number of the boundary lines of the binary image in the whole visual field. This simple method works well for sine-wave and square-wave gratings in our simulations. For more complex background, the number of the boundary also indicates the complexity of the texture to some extent.

\subsubsection{Lamina layer}

Besides the texture estimation, the input image frames are also processed by the lamina layer where the light intensity change, which insects interest more than the intensity itself, are computed to get the primary information of the visual motion\cite{fu2017mimicking}. Denoting $I(x,y,t)$ as the light intensity of pixel (x,y) in time t, then the intensity change P can be expressed as following:
\begin{equation}
P(x,y,t) = I(x,y,t)-I(x,y,t-1).
\end{equation}
\subsubsection{ON and OFF layer}
And then the luminance changes are separated into two pathways\cite{yue2017modeling}. Specifically, the ON pathway deals with light intensity increments; whilst the OFF pathway processes brightness decrements. That is,
\begin{equation}
\begin{split}
P^{ON}(x,y,t) = (P(x,y,t)+|P(x,y,t)|)/2, \\
P^{OFF}(x,y,t) = |(P(x,y,t)-|P(x,y,t)|)|/2.
\end{split}
\end{equation}
\subsubsection{Delay and correlation layer}

Denoting $D^{ON}(x,y,t)$, $D^{OFF}(x,y,t)$ as the output signal of the ON and OFF detectors for horizontal motion and considering a pure time delay of the magnitude $\Delta T$, we have the following expression:
\begin{equation}
\begin{split}
D^{ON}(x,y,t) = P^{ON}(x,y,t-\Delta T)\cdot P^{ON}(x,y+1,t) \\- \alpha P^{ON}(x,y,t)\cdot P^{ON}(x,y+1,t-\Delta T),
\end{split}
\label{eq4}
\end{equation}
where $\alpha$ is chosen from Zanker's paper\cite{Zanker1999} setting as 0.25  forming a partial balanced model. And $D^{OFF}(x,y,t)$ can be expressed similarly.

\subsubsection{Angular velocity decoding layer}

If the input signals are simulated using \eqref{eq1}, then we can get the output of each detector by \eqref{eq4}. And we can combine the outputs of all detectors to get the final response R. In fact, the angular velocity of the background moving is caused by the flying of the insects. The consistency of the background speed helps us simplify the problem.  And the output signals from all ON or OFF detectors in the visual field are averaged to get the response which encodes the angular velocity. And the response R is also averaged over a time period to remove fluctuation caused by oscillatory input.  And the response R is actually a function of the angular velocity and spatial frequency and can be roughly expressed in theoretical \cite{Zanker1999} as the following equation:
\begin{equation}
\begin{split}
R(V, \lambda) \approx & \frac{1-\alpha}{(1+C)^2} + \frac{C^2}{2(1+C)^2}[\sin(\frac{2\pi (\varphi-V \Delta T)}{\lambda}) \\ 
& - \alpha \sin(\frac{2\pi (\varphi+V \Delta T)}{\lambda})].\\
\end{split}
\label{eq5}
\end{equation}

Actually it is hard to derive angular velocity directly from \eqref{eq5}. But we can decode the angular velocity information from the response $R(V,\lambda)$ using an approximation method. Considering that the temporal frequency response curves of the different spatial frequencies have a similar shape, we can use a fitting function $f$ to simulate this shape. Though there is an inevitable fitting error, we can decrease it into an acceptable level if the fitting function is chosen well. One decoding function can be chosen as the following to approximate the actual angular velocity:
\begin{equation}
\widehat{V} = a^*\widehat{\lambda}^{b^*} \sqrt{R},
\end{equation}
where $\widehat{V}$ denotes the decoded angular velocity and $\widehat{\lambda}$ is the estimated spatial period from texture estimation layer. Parameters $a^*$ and $b^*$ can be learned by minimizing the difference from the ground truth using the following equation:
\begin{equation}
(a^*,b^*) = \arg \min_{a,b}\ (V - a\lambda^b \sqrt{R(V,\lambda)}).
\end{equation}
\subsection{Tunnel Centring using AVDM}
Having embedded the AVDM into a virtual bee, first we aimed to investigate the performance of the tunnel centring experiments to see if the virtual bee can centre itself in a narrow tunnel as the real honeybee does \cite{Sri1996}. Secondly, we need to check whether the virtual bee can adjust its position by balancing the image velocities on both eyes when one of the wall is moving in the tunnel experiment \cite{Sri1997}. Here, the angular velocities of the image motion on both eyes are estimated separately. For simplicity, we mainly consider the lateral other than the frontal vision fields. And we assume that the orientation of the head is roughly parallel to the central path of the tunnel and is seldom affected by the body movement. In fact, this can be achieved by gaze stabilization which use the head yaw turn ahead of the body yaw turn to against rotation \cite{boeddeker2010fine}.

In previous section, the angular velocity is measured by the angular displacement $\Delta \phi$ in a small time interval $\Delta t$, that is $V = \frac{\Delta \phi}{\Delta t}$. In the tunnel experiments, the angular velocity can also be defined as the relative tangential velocity of an insect toward the object divided by the distance to it, representing the speed of the object or background moving across the retina. Denoting $v_{bee}$ as  the forward flight speed of the virtual bee and $d$ as the distance to the left wall which moves at a speed $v_{wall}$, the angular velocity of the image motion in its left eye $V_L$ can also be expressed as
\begin{equation}
V_L = \frac{|v_{bee}- v_{wall}|}{d},
\label{tc1}
\end{equation}
The angular velocity of the image motion in the right eye $V_R$ can be similarly derived. This can help us to set up a reasonable simulation environment and provides a reference for the angular velocity estimation. However, in our simulation experiments, AVDM does not need these additional information but only the image frames received by the two simulated compound eyes. The position of the virtual bee in the tunnel is adjusted by balancing $V_L$ and $V_R$ estimated in two eyes. The control scheme for the tunnel centring is shown in the following algorithm, where function $Sign(x)$ is defined as:
\begin{equation}
Sign(x) = \left\{
\begin{aligned}
1 & & if\ x > 0, \\
0 & & if\ x = 0, \\
-1 & & if\ x <0.
\end{aligned}
\right.
\end{equation}

\begin{algorithm}[htb]
\label{algA}
  \caption{Tunnel centring algorithm}
  \KwIn{initial distance to left wall $d_0$, initial distance to entrance $x_0$,  integer history size $m = 10$, iteration index $k =1$;}
  \KwOut{the trajectory of the virtual bee $T = ((d_0, x_0), (d_1, x_1), ..., (d_n, x_n))$}

  \While{ $k < m$}{
       Receive Image from left $I_L(k)$ and Image from right $I_R(k)$\;
       Update $d_{k+1} = d_k$, $x_{k+1} = x_k + \Delta x$\;
  }
  \textbf{end} \\
  \While{ $m \leq k < n$}
    {

      Receive Image $I_L(k)$ and calculate angular velocity $V_L(k)$ using AVDM\;
      Receive Image $I_R(k)$ and calculate angular velocity $V_R(k)$ using AVDM\;

      Update $d_{k+1} = d_k + \Delta d * Sign(V_L(k)-V_R(k))$, $x_{k+1} = x_k + \Delta x$, k = k+1\;
      Discard image frame $I_L(k-m), I_R(k-m)$ from memory storage\;
    }
    \textbf{end} \\
    Return the trajectory T\;
\end{algorithm}

The parameters of the proposed model are shown in the TABLE \ref{table1}. The parameters of the input signal simulation is based on the spatial and temporal processing ability of the honeybee as mentioned above. Other parameters are tuned manually based on our empirical knowledge and will not be changed in the following simulations.
\begin{table}[htb]
\caption{Parameters of the angular velocity decoding model}
\centering
\begin{tabular}{c c c c}
\hline
Eq.  &   &   &  Parameters      \\ \hline
(1)  &   &   &  $\varphi = 2^\circ$, C =1      \\
(5)  &   &   &  $\Delta T = 20  $ ms,  $\alpha = 0.25$       \\
(7)  &   &   &  $a^* = 100$, $b^* =1$ \\
(9)  &   &   &  $v_{bee} = 0.35 $ m/s, $v_{wall} = \pm 0.1 $ m/s \\
\hline
\end{tabular}
\label{table1}
\end{table}

\section{Experiments and Results}
Within this section, we present the experiments and results. The proposed model was first  tested by synthetic grating stimuli to show its spatial independence in Matlab ($\copyright$ The MathWorks, Inc.). And then the model is implemented on a virtual bee using Unity ($\copyright$ Unity Technologies) to simulate the tunnel centring behaviours of the honeybees.

\subsection{Angular velocity decoding results}
In the first kind of experiments, we aimed to inspect the proposed spatial frequency independence of this angular velocity decoding model. Therefore, we used a wide range of spatial period of the grating stimuli as the input signals. The input signals of the moving frames are processed by the proposed model to give an estimation of the angular velocity. In order to get a more general results, the spatial period of the grating is chosen widely from 12$^\circ$ to 72$^\circ$ to show the spatial frequency independence of the decoding model.

\begin{figure}[htb]
\centering
\includegraphics [width=66mm]{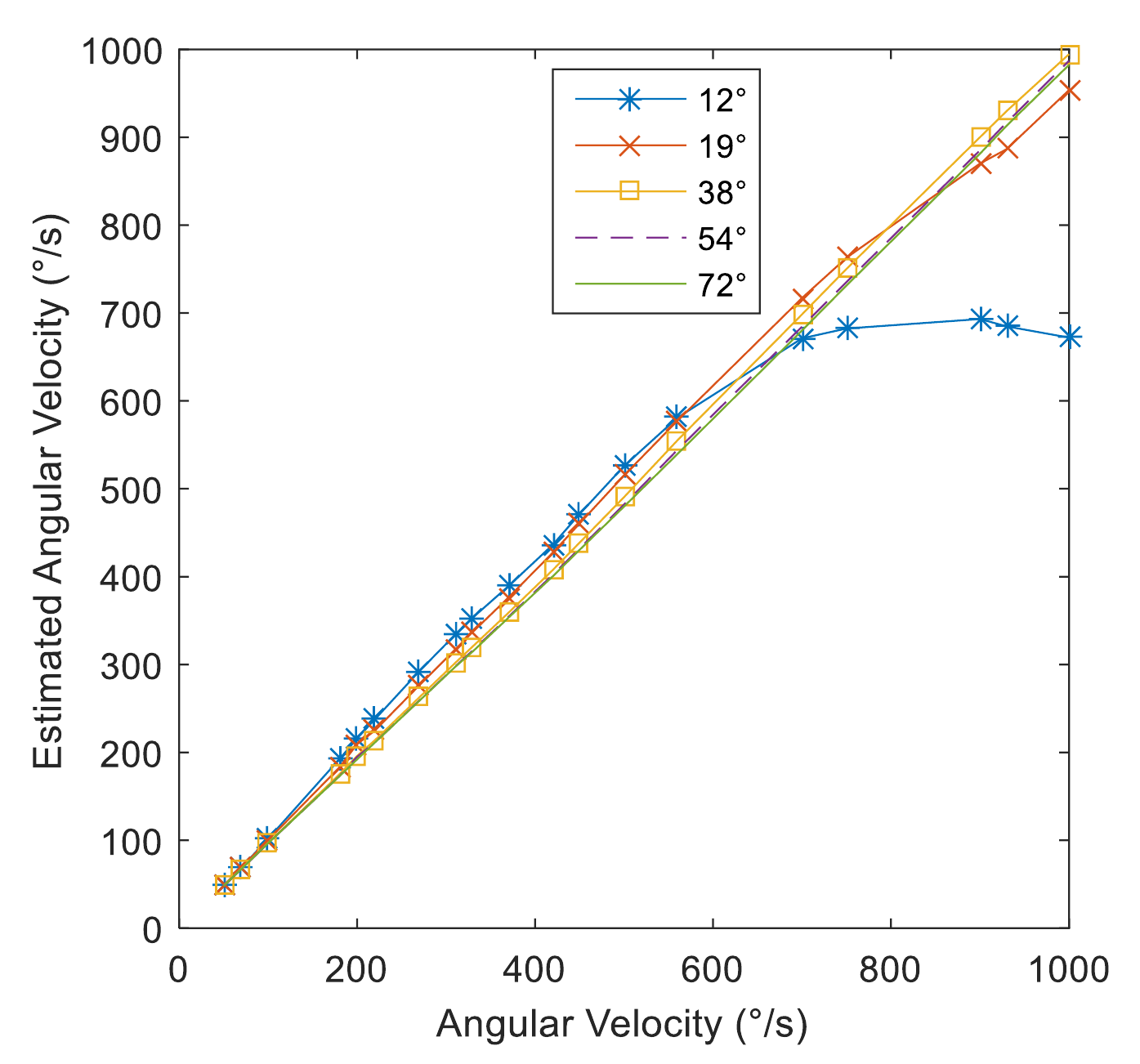}
\caption{ The estimated angular velocity curves from decoding under different angular velocities when tested by moving gratings of different spatial periods (12$^\circ$, 19$^\circ$, 38$^\circ$, 54$^\circ$ and 72$^\circ$).}
\label{fig4}
\end{figure}

The results demonstrate the proposed model has expected independence to spatial frequency of the image motion, see Fig. \ref{fig4}. The angular velocities under different spatial periods are well decoded with little variance expect when the grating is too narrow (12$^\circ$). This is caused by the much higher temporal frequency when the angular velocity is larger than 700$^\circ/s$ for grating of 12$^\circ$. Actually it does not affect the honeybee's flight in most of the case since honeybees tend to maintain a constant angular velocity of 300$^\circ/s$ \cite{Baird2005}, around which our model shows pretty enough spatial independence.

We also used the adjusted R-squared method to analyze the biases from the ground truth for the curves of different spatial periods. The adjusted R-squared values for different spatial periods are computed to see how well the model decodes the angular velocity , see TABLE \ref{table2}. Most of the decoding curves of different spatial periods estimate the angular velocity very well since the adjusted R-squared values are very close to 1. This means though the spatial period changes a lot, the estimated angular velocity varies a little. This ensures that honeybees can estimate the angular velocity accurately when the texture of the background changes in an open flight.

\begin{table}[htb]
\caption{The adjusted R-squared values of the angular velocity decoding curves of different spatial periods. }
\centering
\begin{tabular}{|c|c|c|c|c|c|c|c|}
\hline
Spatial Period      & 12$^\circ$    & 19$^\circ$    & 38$^\circ$     & 54$^\circ$      & 72$^\circ$     \\ \hline
Adjusted-$R^2$  & 0.8685    & 0.9962     & 0.9995     & 0.9981  &0.9974 \\ \hline
\end{tabular}
\label{table2}
\end{table}

\subsection{Contrast of two other angular velocity detecting models}
\begin{figure}[htb]
\centering
\includegraphics [width=63mm]{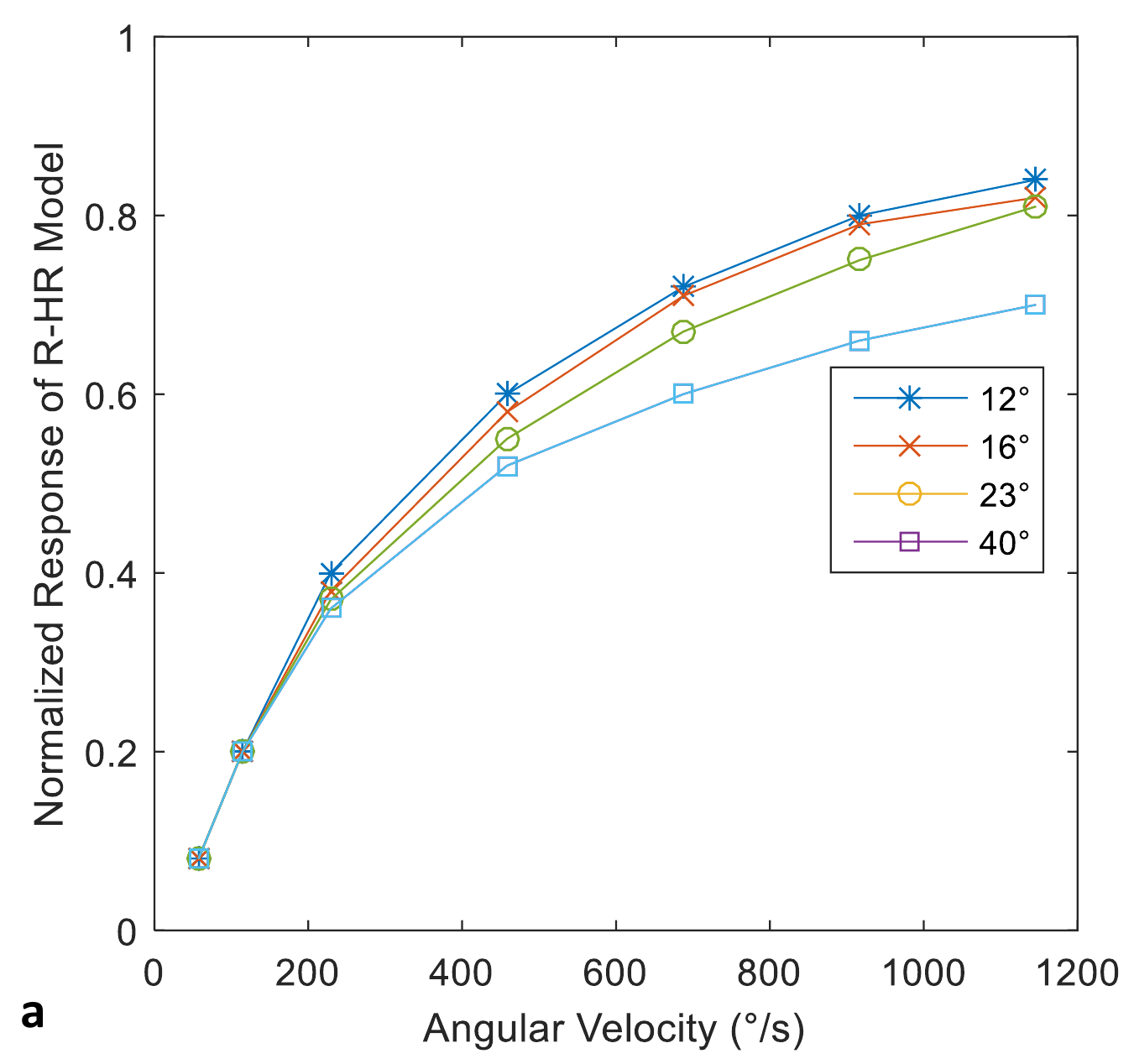}
\includegraphics [width=63mm]{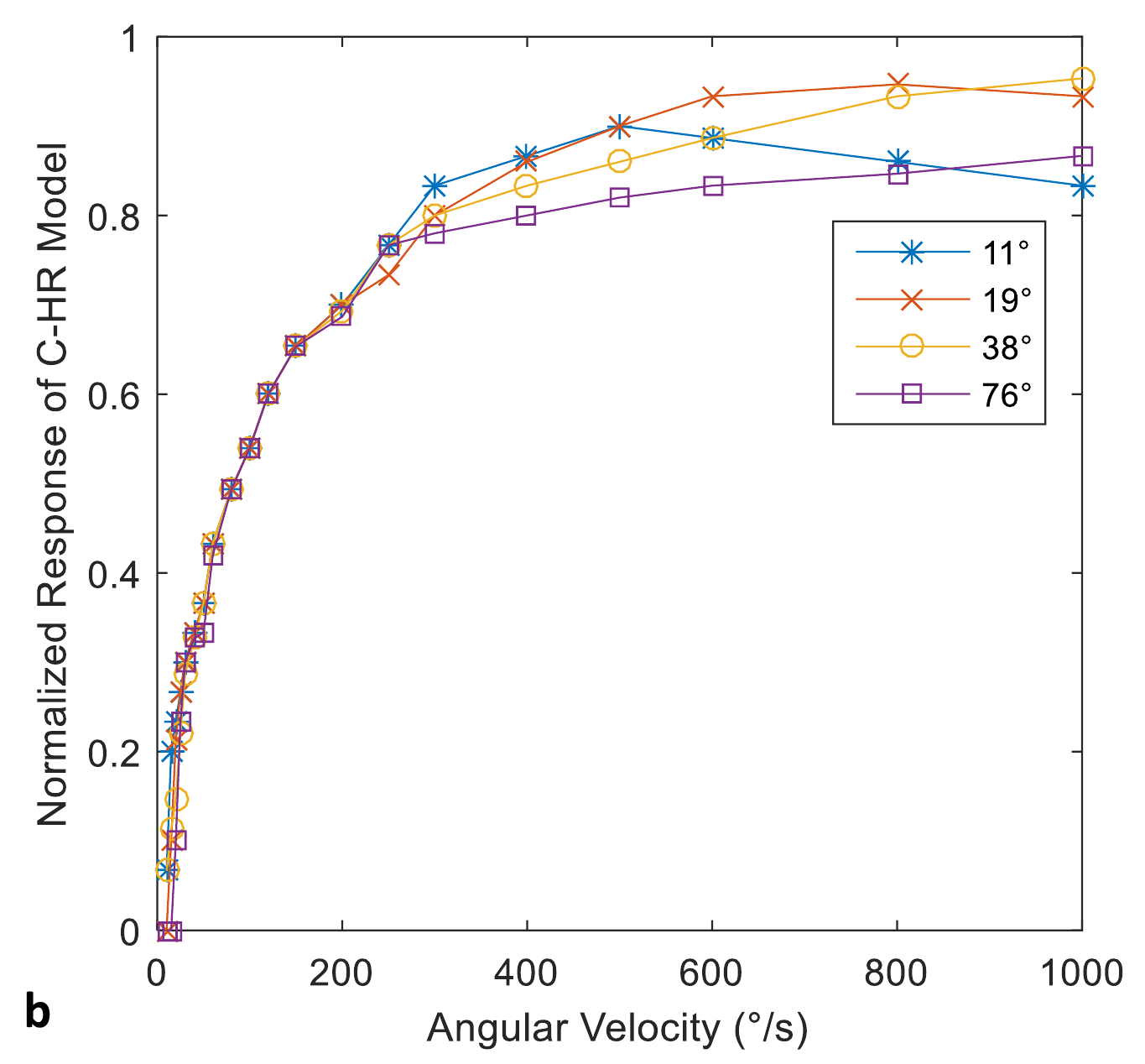}
\captionsetup{font={small},name={Fig.}}
\caption{Contrast of the responses of two other models. (a) Response curves of the R-HR model for different spatial periods show roughly spatial independence\cite{Ria2009}. (b) C-HR model shows largely spatial independency less than the velocity of 200 $^\circ/s$\cite{Cope2016}.}
\label{fig5}
\end{figure}

In order to show that the proposed model has a better spatial independence, we contrast our model with two other detecting models we mentioned above, R-HR model \cite{Ria2009} and C-HR model \cite{Cope2016}. The original results of their models are reploted in Fig. \ref{fig5} under the same metric for a better comparison.

In general, our model show a much stronger spatial independence than two other models, and this independence shows less difference when the angular velocity varies while C-HR model only performs well around 100$^\circ/s$ and R-HR model shows larger difference when angular velocity increases. We should also mention that the original result of the C-HR model use spikes as the output which makes the response curves converge more easily since every detect unit inside the model can only have an integral output. That means it is hard for C-HR model to discriminate approximate angular velocities.

\subsection{Tunnel centring simulation results}

\begin{figure}[htb]
\centering
\includegraphics [width=50mm]{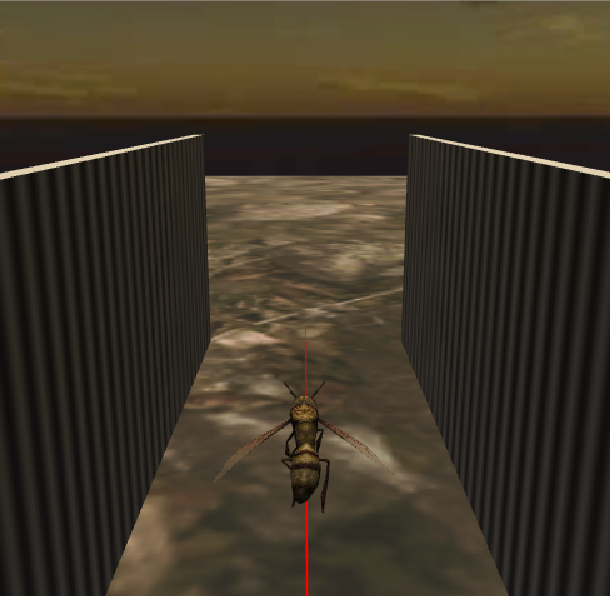}
\captionsetup{font={small},name={Fig.}}
\caption{ Unity simulation environment of the tunnel experiments. Using Unity engine, the images received by two eyes can be processed separately in real time to deicide the route of the flight. A demo video can be found at https://youtu.be/QXl95E71cTE.}
\label{fig6}
\end{figure}

We designed a series of tunnel centring simulations to verify the effectiveness of our angular velocity decoding model. In the tunnel experiments, the virtual bee flies in a 120 cm long, 20 cm wide and 20 cm high simulated tunnel with sinusoidal patterns on both walls (see Fig. \ref{fig6}).

In the real biological behavioural experiments, honeybees can fly in the central of the tunnel with patterned walls of different spatial frequencies. What's more, the honeybees will move towards one side if the wall moves in the same direction of the flight, and move towards the opposite side if the wall moves in the opposite direction of the flight. This means that honeybees can estimate the background speed independent of the spatial frequency and adjust its position by balancing the angular velocity sensed by two compound eyes. Our angular velocity decoding model has the similar characteristics and can be used to simulate this behaviour in the tunnel experiments. If the simulation works well, then it can verify the effectiveness and practicability our proposed model.

\begin{figure}[htb]
\centering
\includegraphics [width=83mm]{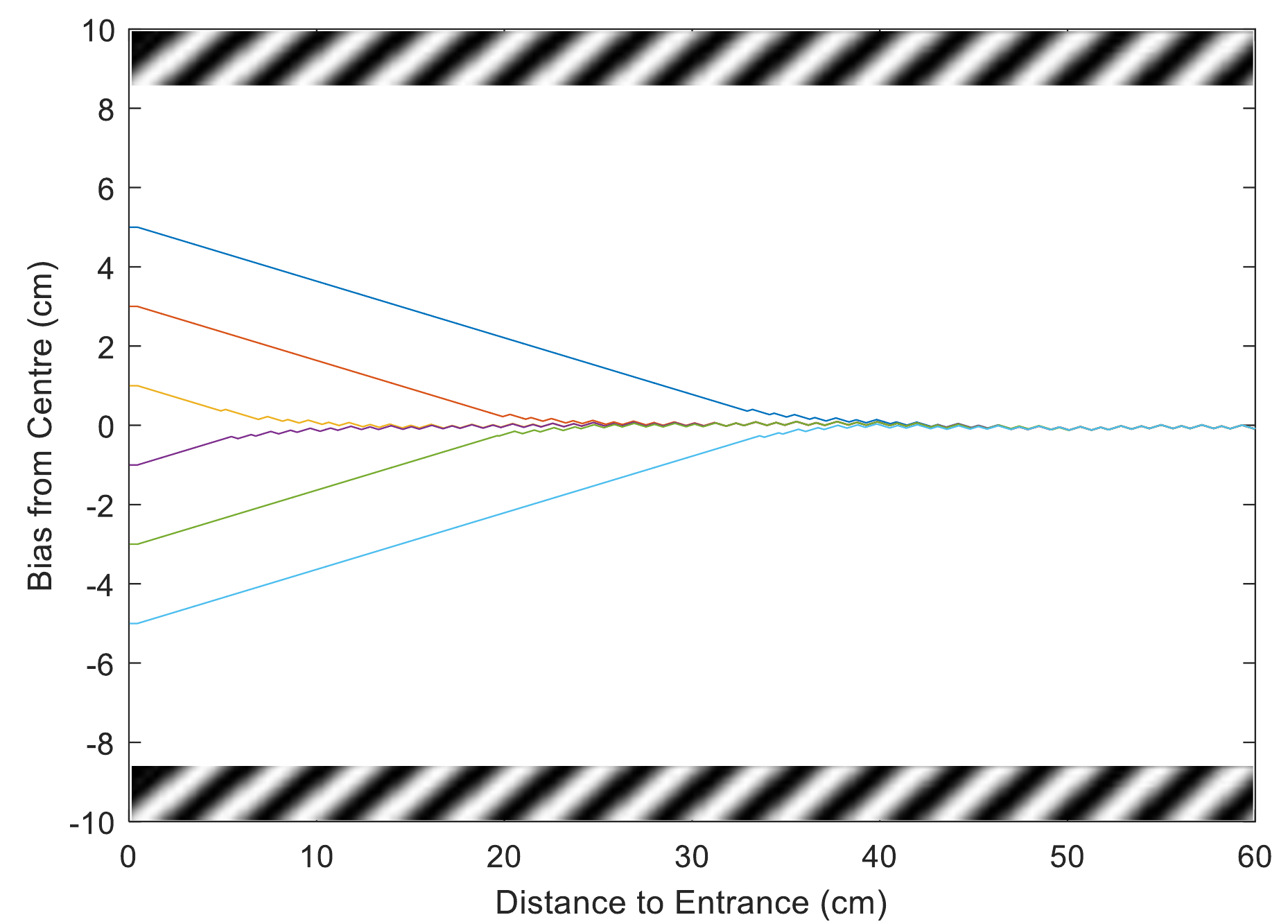}
\captionsetup{font={small},name={Fig.}}
\caption{ Tunnel centring simulations. The flight trajectories of the virtual bee equipped with proposed angular velocity decoding models on both eyes are recorded when released at different start points. All lines of different colors converge to the central path of the tunnel. }
\label{fig7}
\end{figure}

In the first simulation, the virtual bee was released at different start points in the tunnel. The trajectories are recorded to see if it can adjusts its position in the tunnel by balancing the angular velocity estimated on both eyes and finally flies in the central of the tunnel. We implemented the angular velocity decoding model into the eyes of the virtual bee, and then test if the agent can perform a centring response as the real bees. The path recording of the simulations when the two walls carry patterns of same spatial frequency shows in Fig. \ref{fig7}.

As you can see, though the virtual honeybee was released at different start points, it can adjust its position and finally fly following the central path of the tunnel. The result maintains the same if the spatial frequencies are changed (23, 35, 46, 69 $m^{-1}$) as long as the two walls carry the same pattern. The virtual bee will fly through the tunnel with a small bias distance from the center if the walls carry patterns with different spatial frequencies. The biases may be caused by the little difference of the estimated angular velocity when tested by different patterns (see Fig. \ref{fig4}). This means the model is not fully spatial independent. But the real bumblebee behavioural experiments reveals that similar phenomenon can be observed in this situation \cite{Dyhr2010}, which indicates large rather than full spatial independence implemented in the real honeybee's neural system. Further researches are designed to compare the biases between behavioural experiments and tunnel simulations.

In the second simulation, the virtual bee was released at the central of the tunnel. And one of the patterned walls is moving at a constant speed along the flight direction or in the opposite direction to see if the virtual bee can adjust its position as the real honeybees. The simulation results are shown below.

\begin{figure}[ht]
\centering
\includegraphics [width=77mm]{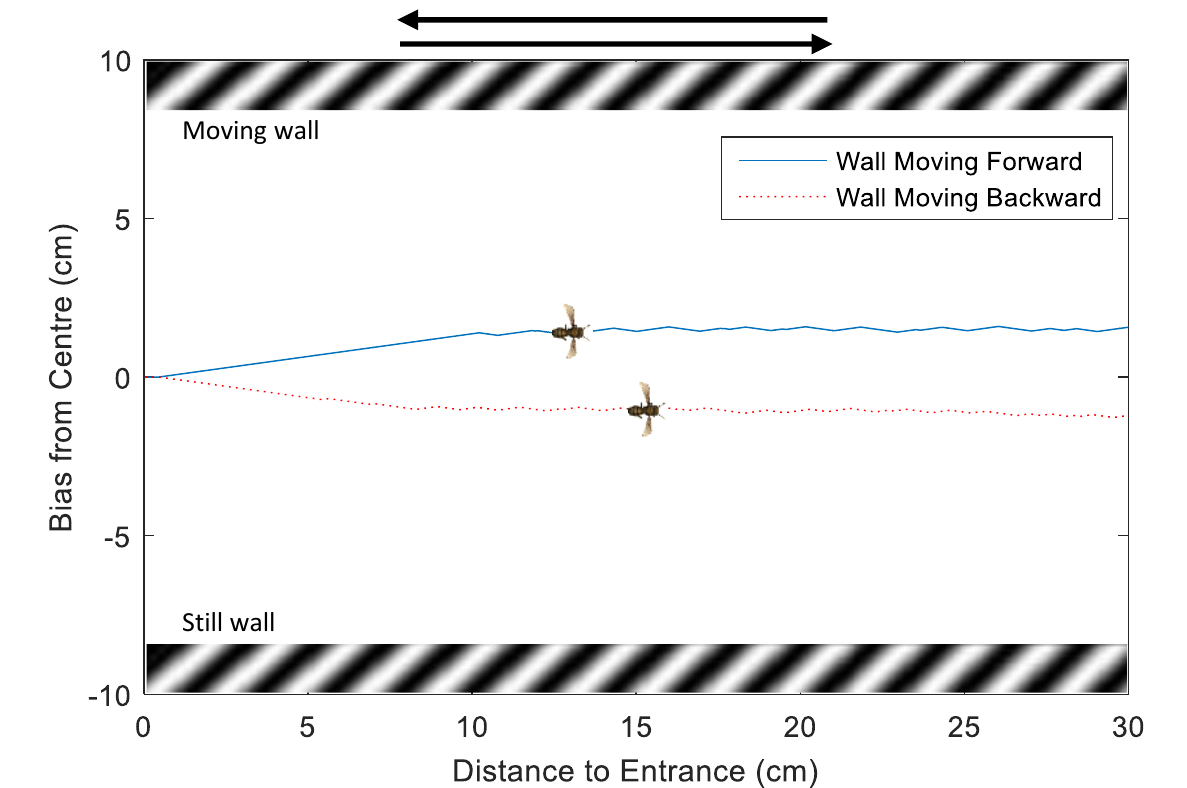}
\captionsetup{font={small},name={Fig.}}
\caption{Tunnel centring with one wall moving forward and backward. The blue solid line indicates flight trajectory when left wall is moving forward and the red line indicates the honeybee's moving trajectory. The virtual bee flying in the patterned tunnel can adjust its position if one of the walls is moving forward or backward at a small constant speed.}
\label{fig8}
\end{figure}

 The virtual bee will move closer to left wall if the left wall is moving along the flight direction at a constant speed (much slower than the flight speed). This is because the angular velocity estimated on left eye is smaller than it was when the wall starts moving forward. Thereby the trajectory shifts towards to the moving wall to balance the angular velocities estimated on both eyes. On the contrary, the trajectory of the virtual bee shifts to right wall if the left wall is moving backward (Fig. \ref{fig8}). Both coincide with the behavioural experiments of the honeybee's visual control \cite{Sri1997}, indicating that the proposed model can explain the behaviours of the honeybee very well.

\section{Conclusion and Discussion}

We have proposed a bio-plausible model, the angular velocity decoding model, for estimating the image motion velocity combining both spatial and temporal information from visual input signals. Most importantly, the model shows large independence of the spatial frequency when tested by moving gratings with a wide range of spatial frequencies. And the model has been implemented in the simulations of a tunnel crossing scenario to reproduce centring response similarly to the honeybee in the tunnel experiments.

Furthermore, the proposed model has great potential to simulate more behaviours of the honeybees. First, the proposed model can be used in the honeybee's speed regulation when the width of the tunnel changes, since honeybees tend to maintain a constant angular velocity (around 300$^\circ/s$) when flying through a tunnel \cite{Baird2005}. Our model can help the virtual bee adjust its flight speed so that it flies faster when the width of the tunnel gets wider and vice versa. Similarly, our model can also be used in the landing simulations like the real honeybees which smartly decreases the flight speed by maintaining a constant angular velocity when getting closer to the ground or landing target \cite{Sri2000}. In addition, this model can be used as the fundamental part of the visual odometer by integrating the angular velocity decoded, which may also provides a possible explanation about how honeybees estimate flight distance. All these simulation experiments will be conducted in the near future to show the effectiveness of our model.

Another significance of this research is that we modelled the ON and OFF pathways in the angular velocity decoding model. According to our calculation, the response curves of the ON and OFF pathways are very similar in most of the simulations. This might be caused by the regularity of the input sinusoidal gratings. The importance of the separation of the ON and OFF pathways may show up when flying in a cluttered environment. In future work, we will investigate the way to combine the ON and OFF signals either averaging over two pathways or obeying a winner-take-all law, expecting to improve the robustness of the model in more complex and cluttered environments. In addition, motion detectors for upward and downward can be constructed similarity and the visual signals are processed separately for each specific direction in parallel. In general, we aim to improve the flexibility of the proposed model adaptive to more complex and dynamic visual scenes. We will also investigate the potential applications in mobile machines like robots and UAVs.

\bibliographystyle{IEEEtran}

\bibliography{IEEEabrv,Reference}

%

%

\end{document}